# Congestionamento Aeroportuário, Escassez de Capacidade e Planejamento na Macrometrópole Paulista

Thayla M. G. Iglesias
Alessandro V. M. Oliveira
Instituto Tecnológico de Aeronáutica
Instituto Tecnológico de Aeronáutica. Praça Marechal Eduardo Gomes, 50. 12.280-250 - São José dos Campos, SP - Brasil.
E-mail: alessandro@ita.br.

*Resumo*: Este artigo apresenta uma narração analítica sobre os limites de capacidade e os desafios operacionais dos principais aeroportos da Macrometrópole Paulista. A partir de exemplos internacionais, como Londres Heathrow, discute como grandes hubs combinam elevada geração de tráfego com severas restrições físicas, evidenciando como a saturação intensifica atrasos, custos operacionais e pressões por expansão. Analisa a escassez de capacidade como problema econômico central, em que pistas, pátios, portões de embarque e terminais tornam-se recursos críticos cujo uso requer mecanismos administrativos e de mercado, incluindo coordenação de slots, regras de priorização e incentivos regulatórios. Discute as limitações impostas pelos elevados custos de terraplanagem, pelos impactos ambientais e pelos custos de desapropriação, que restringem a expansão física de aeroportos centrais como Congonhas e aumentam a dependência de ganhos de eficiência. Projeções de demanda indicam que a capacidade combinada de Congonhas, Guarulhos e Viracopos tende a ser superada, mesmo em cenários conservadores, reforçando a urgência de planejamento integrado. Avaliam-se os efeitos de restrições regulatórias sobre Congonhas, os desafios de ampliação de Guarulhos e as dificuldades estruturais de Viracopos, destacando a relevância estratégica deste sistema multi-aeroportos para sustentar a conectividade nacional. Este artigo foi extraído do capítulo "Um gigante na camisa de força" do livro Voar é para muitos, os negócios das companhias aéreas e a popularização das viagens de avião no Brasil (Edusp, 2023).
*Palavras-chave*: transporte aéreo, planejamento aeroportuário, congestionamento, capacidade, sistemas multi-aeroportos.

## I. Dinâmica estrutural de mega hubs congestionados

O mundo da aviação é imponente. A começar pelos próprios aviões, cada vez maiores, passando pelos terminais aeroportuários gigantes, e finalmente às cifras bilionárias envolvidas nessa indústria. Tudo tem ficado cada vez mais grandioso com o passar das décadas. Mas essa grandiosidade não impede que aviões lotem e que pistas congestionem. Na outra ponta, também possível encontrar casos de empresas aéreas que quebraram, e de aeroportos onde a ociosidade é regra.

E o que há de similar entre 2021, segundo ano de pandemia da COVID-19, e 1966, ano que os Beatles deixaram de fazer apresentações ao vivo? Foi o tráfego no aeroporto londrino de Heathrow, que em fevereiro de 2021 atingiu o pior movimento de passageiros em mais de cinquenta anos, por conta da crise, regredindo aos níveis de 1966[1], quando aqueles rapazes de Liverpool passaram a se dedicar exclusivamente à produção musical em estúdio.

As perdas do Aeroporto de Londres com a crise da COVID-19 chegaram à casa dos bilhões de libras, o que fez com que sua administração requentasse uma proposta de tarifa de congestionamento de 2019, lançada em 2021 como forma de "*proteger o negócio financeiramente e poupar empregos no curto prazo*"[2]. A taxa, entretanto, é bastante inusitada e controversa: aplica-se a carros e táxis que chegam ao aeroporto trazendo passageiros. Olhando em perspectiva, a medida soa como uma iniciativa extrema, mas talvez uma das poucas passíveis de implementação rápida para enfrentar eventos tão negativos.

Em tempos normais, o Aeroporto Londres/Heathrow, que é situado a oeste da capital londrina a pouco mais de 25 quilômetros de distância do centro da cidade, é um exemplo magnífico da pujança da indústria do transporte aéreo. O aeroporto é um dos mais movimentados do mundo, com oitenta milhões de passageiros embarcados e desembarcados em 2018[3]. São mais de mil e trezentos voos por dia, em média. Só em seu Terminal 5, foram mais de 30 milhões de passageiros, e de 200 mil voos naquele ano. É inegável o seu caráter de potência na geração de viagens, negócios e empregos.

---

[1] "Movimento no maior aeroporto de Londres foi o pior em mais de 50 anos", Aeromagazine (15/3/2021), disponível em aeromagazine.uol.com.br.

[2] "Heathrow Airport plan to charge passengers £5 just be dropped off at the terminals", The Sun (3/12/20), disponível em www.thesun.co.uk; "Heathrow Airport plan to charge passengers £5 just be dropped off at the terminals", The Times (23/9/19), disponível em www.thetimes.co.uk.

O aeroporto, entretanto, tem que constantemente lidar com a sobrecarga de demanda e suas consequências. Filas e espera, atrasos e cancelamentos, geração de ruído aeronáutico e emissões de poluentes estão na lista das dificuldades. Problemas que seriam normais em um aeroporto de médio porte, amplificam-se de maneira brutal nesse gigante. Há tempos as autoridades inglesas tentam lidar com o problema, que vem se agravando ano a ano e só arrefeceu com a pandemia. Em fevereiro de 2019, uma nota técnica da consultoria RBB Economics, e disponível no website da Autoridade da Aviação Civil inglesa, a CAA, afirmava que "*o congestionamento do Aeroporto Londres/Heathrow já é um fato bem estabelecido*"[4]. Na Europa continental, a organização internacional Eurocontrol até criou a expressão "aeroportos tipo-Heathrow" para designar situações indesejáveis de operações aeroportuárias com uso de 80% ou mais de capacidade, ou por mais de seis horas consecutivas em um dia[5].

Parece que Londres/Heathrow tornou-se, ao mesmo tempo, um modelo a seguir, em termos de geração de demanda, mas também um exemplo a ser evitado.

A realidade do congestionamento em outros países é similar à do aeroporto londrino em muitos casos. Nos Estados Unidos, a Administração Federal de Aviação, FAA, até antes da pandemia apontava como estando em situação de capacidade limitada os aeroportos de Nova Iorque (todos), Washington/Ronald Reagan e Boston, na Costa Leste, e de Seattle-Tacoma, São Francisco/Internacional e de Los Angeles/Internacional, na Costa Oeste. Outros, como os aeroportos de Charlotte-Douglas e de Las Vegas, também chegariam a essa situação até 2030. Na Europa, de acordo com as expansões previstas de capacidade aeroportuária, a Eurocontrol projetava que os vinte principais aeroportos operariam próximo de sua capacidade até 2040[6]. Na China, a cidade de Pequim precisou construir um novo aeroporto, o Internacional de Pequim/Daxing, para expandir a capacidade e aliviar as pressões de demanda sobre o aeroporto principal, o Internacional de Pequim/Capital[7].

Pesquisadores do Centro Aeroespacial Alemão (DLR) certa vez apontaram que seis por cento de todos os voos realizados no mundo eram operados em condições de congestionamento[8]. Dentre as medidas possíveis de solução para o problema, os pesquisadores indicaram ações como a construção de novas pistas, a reorganização das operações do tráfego aéreo que possibilitem ganhos de eficiência, a realocação de voos para aeroportos secundários, além da utilização de aeronaves com mais assentos. Essas sugestões configuram em novidade "zero", pois são conhecidas há décadas na indústria. O problema é mesmo de implementação.

Congestionamentos aeroportuários trazem desafios tremendos às autoridades aeronáuticas, que devem reforçar o planejamento e discutir com a sociedade todas as alternativas existentes para a expansão da malha de voos. Implementar soluções é, muitas vezes, complexo e envolve negociações. Procrastinar o equacionamento de problemas pode às vezes ser bom, para não se tomar decisões erradas, ao sabor da conjuntura. Mas também pode ser péssimo, aumentando o custo, ou mesmo inviabilizando muitas possibilidades.

## II. MECANISMOS DE GESTÃO DA ESCASSEZ AEROPORTUÁRIA

Aeroportos, como várias outras infraestruturas básicas, podem ter dificuldade para se expandir. Pode-se dizer que têm que lidar com três "Ms" que podem bloquear sua ampliação: "morro", "mato" e "moradias". Em outras palavras, em muitos casos há que se arcar com custos de obras de terraplanagem ("morro"), custos ambientais ("mato"), e os custos de desapropriação advindos de ocupação regular ou irregular em seu entorno ("moradias"). Esses custos podem ser elevados.

---

[3] Todos os números desse parágrafo foram obtidos em "Facts and figures" Heathrow - Our Company, disponível em www.heathrow.com/company/about-heathrow/facts-and-figures, e extraído em 24/9/2021.

[4] "The effect of congestion at Heathrow Airport" (19/2/2019), disponível em www.caa.co.uk.

[5] "European Aviation in 2040. Challenges of Growth - Annex 4 - Network Congestion", Eurocontrol, 2018, disponível em www.eurocontrol.int.

[6] "European Aviation in 2040. Challenges of Growth - Annex 4 - Network Congestion", Eurocontrol, 2018, disponível em www.eurocontrol.int.

[7] "Direto da China: Conheça o novo Aeroporto Internacional de Pequim Daxing", Melhores Destinos (6/12/2019), disponível em www.melhoresdestinos.com.br.

[8] Gelhausen, Berster & Wilken (2013).

Aeroportos com capacidade limitada apresentam restrições operacionais por conta da escassez relativa de seus componentes. É uma questão de volume de demanda elevado em relação às possibilidades de atendimento, em especial em horários de pico. Podem faltar horários para acomodar mais voos na pista de pousos e decolagem, podem ser insuficientes o número de balcões de check-in, os portões de embarque, as esteiras de bagagem, as pontes de embarque e desembarque, dentre diversos outros recursos essenciais do aeroporto, cuja indisponibilidade no pico pode ser considerada crítica. Pode inclusive faltar área no terminal para a movimentação adequada de passageiros e no pátio de manobras para as aeronaves. Na verdade, basta que um desses itens esteja sob pressão para os que os problemas comecem a surgir. Na impossibilidade de expansão, há que se gerenciar, e muito bem, essa escassez.

Internacionalmente, a questão das operações aeroportuárias são alvo de muito estudo. Os reconhecidos papas da área são os professores Robert E. Caves e Norman J. Ashford, da Universidade de Loughborough, Inglaterra, e Robert Horonjeff, da Universidade de Califórnia, em Berkeley[9]. No Brasil, o professor Cláudio Jorge Pinto Alves, do ITA, é unanimidade dentre os especialistas[10]. Não por falta de conhecimento acumulado na área de engenharia padeceremos de problemas em nossos aeroportos.

Se há congestionamento, cada instante de uso de pista de pousos e decolagens, pátio de manobras ou terminal de passageiros se torna crucial. Todas as companhias aéreas presentes no aeroporto comercializaram seus assentos e necessitam utilizar esses recursos aeroportuários dentro do que foi contratado por seus clientes. Na pista, cada formação de fila de aeronaves pode gerar impactos operacionais gritantes, que provocam uma cascata de problemas aos próximos voos. Nessas situações, o uso da pista pelas companhias aéreas passa a ser então mais rigorosamente administrado, ou mesmo regulado, definindo-se horários rígidos denominados "slots".

Segundo a Associação Internacional de Transporte Aéreo (IATA), até antes da pandemia mais de 1,5 bilhão de passageiros anuais, ou 43% do tráfego global, partia de mais de 200 aeroportos coordenados com slots[11].

Recentemente, um estudo da professora Rosário Macário, do Instituto Superior Técnico de Portugal, fez um levantamento das atuais práticas e necessidades futuras na área de planejamento de capacidade, congestionamento, e gestão de demanda em aeroportos[12]. Segundo a pesquisa, a capacidade aeroportuária pode ser alocada de diferentes maneiras, sendo elencados 1. o gerenciamento administrativo, por meio de regras de alocação de slots e distribuição de tráfego; 2. o gerenciamento com base no mercado com leilões de slots, na negociação secundária e na taxação do movimento durante o pico de operações; 3. uma combinação de ambos os mecanismos; e, por fim, 4. ausência de regulamentação. Na Europa, os aeroportos europeus se valem dos mecanismos administrativos, enquanto que nos Estados Unidos, aplica-se a regra de ordem de chegada, com exceção dos cinco mais congestionados. De qualquer maneira, lembra o estudo, o nível de atrasos de voos na Europa é significativamente menor do que no país do outro lado do Atlântico.

O professor Mikio Takebayashi da Universidade de Kobe, no Japão, também investigou formas de fazer a gestão aeroportuária[13]. Ele analisou a situação de áreas de múltiplos aeroportos - aquela região metropolitana ou uma extensão dela, em que existe a opção de escolha de voos de mais de um aeroporto. O professor constatou que a melhor estratégia para as companhias aéreas é operar apenas voos curtos quando se tratar de um aeroporto com restrições operacionais localizado em uma região central da cidade. Em contrapartida, um aeroporto localizado na região periférica da metrópole estaria apto a comportar voos tanto de curta quanto de longa duração.

Um bom exemplo para os resultados obtidos pelo professor são os aeroportos de São Paulo, Congonhas e Guarulhos, em que o primeiro é central e recebe majoritariamente voos domésticos, enquanto que o segundo é periférico comparado ao centro da capital paulistana, e recebe voos domésticos e internacionais.

No entanto, Takebayashi alerta sobre o risco de congestionamentos se o governo não estabelecer diretrizes para as alocações de voos, em que poderá haver uma inversão indesejada na distribuição dos voos entre aeroportos centrais e periféricos: o aeroporto central tenderá a operar com voos curtos e longos, e o periférico apenas com voos curtos, o que, segundo o autor, não geraria benefícios nem para os passageiros, nem para as companhias aéreas.

---

[9] Kazda & Caves (2007) Horonjeff et al. (2010).
[10] Vide módulos e aulas do professor para download em www.civil.ita.br/~claudioj.
[11] "Worldwide Airport Slots", IATA, disponível em www.iata.org.
[12] Cavusoglu e Macário (2021).
[13] Takebayashi (2012).

## III. PROJEÇÕES DE DEMANDA E URGÊNCIA DE CAPACIDADE

Um dos maiores desafios de planejamento de transportes no Brasil diz respeito ao que pode ser chamado de "Questão Aeroportuária Paulistana". Trata-se simplesmente de prover voos à décima mais populosa aglomeração urbana do mundo, a Macrometrópole Paulista. Essa área engloba as regiões metropolitanas de São Paulo, Campinas, Sorocaba, Baixada Santista e São José dos Campos. Se computadas as respectivas regiões geográficas intermediárias do IBGE (mesorregiões), estamos falando de mais de 30 milhões de pessoas.

Em 2019, essa Macrometrópole produziu em torno de 77 milhões de passageiros aéreos. São 211 mil passageiros diários, embarcando e desembarcando em terminais da região, sendo 56% no Aeroporto de São Paulo/Guarulhos, 30% no Aeroporto de São Paulo/Congonhas e 14% no Aeroporto de Campinas/Viracopos. Atualmente, os aeroportos de Congonhas, Guarulhos e Viracopos possuem capacidade para, respectivamente, 17, 50, e 25 milhões de passageiros por ano, o que totaliza 92 milhões.

Projeções de demanda sempre erram, muitos diriam. Entretanto, pode-se enxergar a questão sob uma ótica diferente: previsões de demanda sempre acertam, a realidade é que se antecipa ou retarda a acontecer.

Não é irrealista dizer que, passada a pandemia, muitas previsões de demanda por viagens aéreas indicarão que a movimentação de passageiros na Macrometrópole Paulista irá ultrapassar a capacidade instalada de seus aeroportos em muito breve.

Se fizéssemos uma dessas previsões de demanda, imaginando um crescimento a taxas mais otimistas de 5% ao ano, essa capacidade seria atingida já em meados dos anos 2020. Se fôssemos bem pessimistas, talvez utilizando uma taxa de crescimento de 2,5% ao ano, esse gargalo chegaria mais para o final da década, talvez em 2030.

Na área de previsão, importa a maneira como enxergamos as perspectivas para o mundo, e a conjuntura acaba influenciando muito nisso. Uma crise nos deixa pessimistas. Uma retomada econômica nos deixa otimistas. Dada essa bipolaridade causada pela conjuntura brasileira, sempre é prudente considerar um intervalo de possibilidades entre o cenário mais pessimista e o mais otimista, além de um cenário neutro intermediário.

Mas qualquer que seja a previsão, tudo indica ser mais do que necessário adicionar capacidade aeroportuária extra à existente na Macrometrópole Paulista.

Os números ditam o ritmo da urgência.

## IV. RESTRIÇÕES LOCACIONAIS E EFEITOS REGULATÓRIOS

É fato público e notório que o Aeroporto de São Paulo/Congonhas está localizado em uma região central da capital paulista, rodeado de espaço urbano densamente povoado. Por questões de espaço físico mesmo, não tem como se expandir, encontra-se em uma verdadeira "camisa de força". Por conta disso, o aeroporto perdeu a liderança nacional de movimento de passageiros para o Aeroporto de Guarulhos e, mais recentemente, o Aeroporto de Brasília. Congonhas, entretanto, sempre foi, e continuará sendo, o aeroporto preferido por uma grande parcela dos paulistanos, pela simples proximidade de suas residências e locais de trabalho. Impossível tirar o status de majestade desse charmoso aeroporto que se tornará um "senhor de noventa anos" em pouco tempo.

Dois estudos desenvolvidos pelo Núcleo de Economia do Transporte Aéreo do ITA investigaram os efeitos de restrições operacionais sobre Congonhas, impostas pelas autoridades logo após o trágico acidente aeronáutico de 2007[14].

Na ocasião, o debate sobre os aeroportos de São Paulo ganhou contexto nacional. A crise do "apagão aéreo", em que operações-padrão de controladores de tráfego provocava ondas de atrasos e cancelamentos por todo o país, rapidamente se transformou em uma crise do sistema de aeroportos paulistanos. Nunca o tema se tornou tão falado e discutido, ganhando manchetes de jornais e revistas em uma intensidade que somente foi desbancada posteriormente pelo noticiário da Operação Lava-Jato. Especialistas da área se tornaram verdadeiras "estrelas pop", com aparição frequente na mídia.

O governo, preocupado, cometeu acertos em suas reações ao grave problema da aviação brasileira, mas também deslizou. Um desses deslizes, que desviou bastante a atenção quanto ao cerne da questão, foi a ideia de "fritar" a Diretoria da ANAC, uma agência recém-criada com a finalidade de zelar pela segurança e bem-estar do passageiro, de forma independente[15]. Erro ou acerto, o fato é que os currículos dos indicados à diretoria da agência também se tornaram visivelmente mais técnicos e menos políticos desde então.

---

[14] Iglesias e Oliveira (2015) e Miranda e Oliveira (2018).

Um acerto foi a medida de maior aperto do sistema de slots do aeroporto, cuja disponibilidade caiu de mais de quarenta movimentos horários permitidos à aviação regular, para pouco mais de trinta[16]. Essa medida permitiu trazer uma percepção de maior segurança ao aeroporto, que é, e sempre foi bastante seguro, diga-se de passagem. Atualmente, pode-se dizer que esse patamar de movimentação horária é demasiadamente conservador, entretanto. Errar pra menos, nesse caso, significa perder oportunidades de novas viagens aos passageiros.

Algumas outras medidas adotadas na época para lidar com o congestionamento de Congonhas foram mais polêmicas e de curta duração. Foi aplicada uma vedação de uso do aeroporto para escalas e conexões de voo e também uma "regra de perímetro", onde voos para localidades acima de 1.500 quilômetros de distância não eram mais autorizados.

Os estudos desenvolvidos pelo núcleo de pesquisas focaram justamente esse período de crise, considerado o mais crítico do aeroporto das últimas décadas, talvez de todos os tempos. Os resultados apontam que houve importantes impactos nas operações do aeroporto na ocasião.

Como é de se esperar de restrições operacionais sobre aeroportos, os estudos indicaram que houve tendência à queda de movimento de passageiros, considerando todos os outros fatores dados. Essa queda foi mais relevante para voos com distâncias médias e longas, e se deram no longo prazo, o que em parte explica a perda de liderança do aeroporto em nível nacional. A proporção de passageiros em conexão, de fato caiu, mas apenas no curto prazo. Sabe-se que as companhias aéreas do aeroporto voltaram a praticar as conexões de voo após um certo período, e intensificaram essa estratégia mais recentemente. Quanto a atrasos e cancelamentos, há indicações de que houve um maior controle, e os números não mais saíram dos eixos. Por fim, os pesquisadores encontraram evidências de que a concentração de slots em poucas empresas está associada a um aumento dos preços das passagens aéreas, mas também pode gerar o benefício de uma maior pontualidade dos voos.

Após esse conturbado período, as autoridades responsáveis continuaram à caça de meios de lidar com os problemas do aeroporto. Em particular, buscaram fomentar o desempenho das operadoras na utilização das pistas de pouso e decolagem, bem como a possibilidade de maior oferta de voos, especialmente se advindos de novas companhias aéreas. Assim, em julho de 2014, a ANAC, divulgou novos regulamentos que permitiriam maior acompanhamento, controle e redistribuição mais equânime de slots[17]. A regularidade e a pontualidade dos voos passaram a ser critérios usados nas alocações dos recursos do aeroporto dali em diante. O intuito da agência é incentivar a concorrência e a prática de menores preços de passagens aéreas, mesmo se não houver aumento da capacidade aeroportuária.

Empresas com menor operação no aeroporto, como Azul e Avianca (que posteriormente quebrou), foram beneficiadas com a nova resolução, ampliando a sua participação nas posições do aeroporto. Congonhas, entretanto, ainda é bastante concentrado em voos das duas maiores companhias aéreas do Brasil.

# V. Reflexões sobre o sistema multi-aeroportos paulistano

Desde o início do milênio, em que o tráfego aéreo nacional se intensificou aceleradamente, passando pelas complicações do período de "apagão aéreo" e o trágico acidente aeronáutico no Aeroporto de Congonhas do final dos anos 2000, a Questão Aeroportuária Paulistana está posta às autoridades brasileiras.

Também há algum tempo, mais precisamente desde o início dos anos 2010, o governo federal tomou a decisão de resolver os problemas de capacidade aeroportuária brasileiros buscando investidores na iniciativa privada. A privatização e a regulação de aeroportos se tornaram a tônica maior das políticas públicas do transporte aéreo.

---

[15] "Diretoria da Anac articula pedido de renúncia coletiva", Folha de São Paulo (28/7/2007), disponível em www1.folha.uol.com.br.

[16] As restrições operacionais foram as seguintes: (i) operações restritas a voos direto (ponto a ponto), sem conexões ou escalas de voos; (ii) operações de partidas e chegadas limitadas a distância máxima de 1500 quilômetros (origem-destino) - ambas descritas na Portaria ANAC nº 806/2007, de julho de 2007, sendo a última atualizada conforme Portaria ANAC nº 327/2008 (ANAC, 2007; 2008a). Ainda, os pousos e decolagens foram limitados a 15 pares de slots/hora, em se tratando da Aviação Regular, e 2 pares/hora para a Aviação Geral (ANAC, 2008b).

[17] Resolução ANAC nº 336/2014.

Os aeroportos de Guarulhos e Viracopos foram entregues à administração privada em 2012. Entretanto, Viracopos entrou em recuperação judicial e teve sua concessão devolvida para uma relicitação[18]. Dos grandes aeroportos da Macrometrópole Paulista, apenas Congonhas ainda pairava sob a administração estatal da Infraero. Não mais, no que se seguirá na atual década. A decisão de ter todos aeroportos sob concessão privada foi tomada[19]. As dezenas de milhões de passageiros anuais dessa região serão movimentadas em aeroportos privados que terão de cumprir metas de eficiência e qualidade da prestação de serviços, segundo os atuais regulamentos da Agência Nacional de Aviação Civil, ANAC.

Mas a pergunta que fica é: como se dará a expansão de capacidade desse imenso complexo de terminais que abarca a mais movimentada região de múltiplos aeroportos do Brasil e uma das maiores do mundo? Em seus detalhes, como será equacionada a Questão Aeroportuária Paulistana? Qual é o plano?

Para tratar desse tema, há que se olhar para as demais alternativas existentes, que não Congonhas. Os aeroportos de Guarulhos e Viracopos apresentam os seus desafios próprios. Ambos aeroportos passaram por melhorias e inauguraram novos terminais após a privatização. Ambos expandiram suas capacidades de movimentação de passageiros. Em ambos os casos, um novo ciclo de investimentos em capacidade permitiria lidar com o problema.

Mas para viabilizar um verdadeiro salto de crescimento, esses aeroportos necessitariam de mais uma pista de pousos e decolagens - Viracopos, a segunda, e Guarulhos, a terceira. Ou pelo menos um deles.

Mas o fato é que Campinas não é uma opção tão próxima da cidade de São Paulo, que abrange grande parte das zonas de geração de viagens da região. A pouco menos de cem quilômetros do centro da capital, pode-se imaginar o custo aos passageiros em realizar deslocamentos ao aeroporto. Pense na emissão de gases poluentes advindos dessas viagens. Melhorar o acesso dos passageiros vira então prioridade. Mas talvez isso não seja o suficiente. Seria preciso aumentar muito o seu movimento de forma a disparar algum gatilho que viabilizasse os investimentos para a materialização da segunda pista.

Já Guarulhos é mais próximo da capital, cerca de 25 quilômetros do centro. Mas o aeroporto tem o sério problema da ocupação urbana que se formou irregularmente na sua área reservada à terceira pista. Um projeto de expansão precisaria bancar os custos de desapropriação de famílias das redondezas, ou propor soluções alternativas inovadoras de desenho de aeroportos.

Muitos argumentam que a solução dos problemas aeroportuários de São Paulo passaria pela construção de um novo aeroporto. Outros discordam.

Enfim, propostas de soluções não faltam. Com certeza nenhuma delas é de implementação fácil e qualquer uma delas vai requerer muita vontade política para enfrentar o problema. O advento da crise sanitária amplificou as incertezas, mas também de certa forma proporcionou uma possibilidade de postergação do seu equacionamento. Agora é correr contra o tempo.

Uma coisa é certa: não obstante a pandemia da COVID-19, em algum momento a demanda por voos prosseguirá sua trajetória de ascensão. Isso porque fatores como a renda per capita e a disposição a pagar de paulistanos e paulistas, continuarão materializando o desejo de viajar em efetivo movimento aeroportuário nos terminais aeroportuários da região. Ademais, o restante do país também necessita de alta conectividade com São Paulo, que é centro nacional de referência em negócios, eventos, saúde, educação e turismo.

Sem um acesso rápido e de qualidade a São Paulo, permitido por seus aeroportos, o Brasil como um todo sai no prejuízo.

# Referências

---

[18] "Segunda pista, outorga de R$ 3,5 bilhões e fim de desapropriações: entenda a nova licitação do Aeroporto de Viracopos", G1 (9/9/2021), disponível em g1.globo.com.

[19] "Aeroportos de Congonhas e Santos Dumont são incluídos em programa de privatização. Veja lista", UOL (23/2/2021), disponível em congressoemfoco.uol.com.br.